\documentclass[prd,twocolumn,showpacs,floatfix,amsmath,amssymb,floatfix]{revtex4}
\usepackage{graphicx,color,dcolumn,booktabs,bm}
\usepackage{float}
\usepackage{longtable,lscape}
\usepackage{txfonts}
\usepackage{overpic}
\usepackage{amssymb}
\usepackage{indentfirst}
\usepackage{feynmf}   
\usepackage{slashed}  
\usepackage{cases}
\usepackage{color}
\usepackage{multirow}
\usepackage{epstopdf}
\usepackage{graphicx,color,dcolumn,booktabs,bm}
\def\stat{(\mathrm{stat.})}
\def\syst{(\mathrm{syst.})}
\usepackage[colorlinks,
            citecolor=blue,
            anchorcolor=red,
            menucolor=red,
            linkcolor=red,
            filecolor=red,
            runcolor=red,
            urlcolor=blue,
            frenchlinks=red]{hyperref}
\begin{document}

\title{Strong decays of the $\Lambda_{c}(2910)$ and $\Lambda_{c}(2940)$ in the $ND^{*}$ molecular frame}
\author{Zi-Li Yue$^{1}$}
\author{Quan-Yun Guo$^{1}$}
\author{Dian-Yong Chen$^{1,2}$\footnote{Corresponding author: chendy@seu.edu.cn}} 
\affiliation{
 $^{1}$ School of Physics, Southeast University,  Nanjing 210094, China} 
\affiliation{$^{2}$Lanzhou Center for Theoretical Physics, Lanzhou University, Lanzhou 730000, China}
\begin{abstract}
Stimulated by the observation of a new structure, named $\Lambda_{c}(2910)$, in the $\Sigma_{c}(2455)^{0,++}\pi^{+,-}$ decay channel from B meson decay process by the Belle Collaboration, and the similarity of the $\Lambda_{c}(2910)/\Lambda_{c}(2940)$ and $P_{c}$ states, we investigate the decay behavior of the $\Lambda_{c}(2910)$ and $\Lambda_{c}(2940)$ in the $N D^{*}$ molecular frame with the possible $J^{P}$ quantum numbers to be $1/2^{-}$ and $3/2^{-}$. We employ an effective Lagrangian approach to evaluate the partial widths of $ND$, $\Sigma_{c}\pi$ and $\Sigma_{c}^{*}\pi$ channels. The estimations in the present work indicate that the $J^P$ quantum numbers of $\Lambda_{c}(2910)$ and $\Lambda_{c}(2940)$ are preferred to be $1/2^{-}$ and $3/2^{-}$, respectively. From the present estimations, we also find the branching ratio of $\Lambda_c(3/2) \to \Sigma_c^\ast \pi$ is much larger than that of $\Lambda_c(1/2) \to \Sigma_c^\ast \pi$, thus $\Sigma_{c}^{*}\pi$ could be a good channel to distinguish the $J^{P}$ quantum numbers of $\Lambda_{c}(2910)$ and $\Lambda_{c}(2940)$. Therefore, we suggest searching for the structure in $\Sigma_{c}^{*}\pi$ invariant mass distribution in Belle $\mathrm{\uppercase\expandafter{\romannumeral2}}$.
\end{abstract}


\maketitle

\section{Introduction}
\label{sec:introduction}

In the pentaquark family, the prominent members are the $P_c$ states, which were first observed in the $J/\psi p$ invariant mass distribution of the process $\Lambda_b \to J/\psi p K^-$ by the LHCb Collaboration in 2015~\cite{LHCb:2015yax}. Subsequently, the LHCb Collaboration reanalyzed the same process utilizing the data collected in Run I and Run II in 2019~\cite{LHCb:2019kea}. The new analysis reported a new narrow state, $P_c(4312)$, with a statistical significance of $7.3\sigma$, and a two-peak structure in the previously reported $P_c(4450)$ corresponding to $P_c(4440)$ and $P_c(4457)$, with a statistical significance of $5.4 \sigma$, while the broad $P_c(4380)$ was described by the background line shape~\cite{LHCb:2019kea}. The nature of the $P_c$ triplets is of great interest to theorists, and some exotic interpretations, such as pentaquark \cite{Anisovich:2015cia,Chen:2016otp,Santopinto:2016pkp,Park:2017jbn,Zhu:2015bba,Ghosh:2015xqp,Giannuzzi:2019esi,Wang:2019got} and molecular state \cite{Roca:2015dva,Lu:2016nnt,He:2015cea,Yamaguchi:2016ote,Chen:2015loa,Eides:2015dtr,Xiao:2019mvs,Wu:2019rog,Liu:2019tjn,Zhang:2019xtu,Guo:2019kdc,Guo:2019fdo}, have been proposed.      

In addition to the $P_c$ states, there are some not-so-obvious candidates of pentaquark states, such as $\Lambda_c(2940)$, which was first observed in the $D^0p$ invariant mass distributions by the BABAR Collaboration in 2006~\cite{BaBar:2006itc}. The mass and width were reported to be $(2939.3 \pm 1.3 \stat \pm 1.0 \syst )$ MeV and $(17.2 \pm 1.3\stat \pm 1.0 \syst )$ MeV, respectively. Later on, the Belle Collaboration confirmed the existence of $\Lambda_c(2940)$ in the $\Sigma_{c}^{0,++}\pi^{+,-}$ invariant mass distributions~\cite{Belle:2006xni}. In addition, the LHCb Collaboration performed an amplitude analysis of the decay $\Lambda_b^0 \to D^0 p \pi^-$ by using the data sample corresponding to an integral luminosity of $3.0 ~\mathrm{fb}^{-1}$ of $pp$ collisions~\cite{LHCb:2017jym}. The analysis indicated that the most like spin-parity assignment for $\Lambda_c(2940)$ is $J^P=3/2^-$ but the other solutions with spins $1/2-7/2$ cannot be excluded~\cite{LHCb:2017jym}. So far the PDG average of the mass and width of the $\Lambda_{c}(2940)^{+}$ are~\cite{ParticleDataGroup:2022pth}
\begin{eqnarray}
m_{\Lambda_{c}(2940)^{+}}&=& (2939.6^{+1.3}_{-1.5})~\mathrm{MeV},\nonumber\\
\Gamma_{\Lambda_{c}(2940)^{+}}&=&(20^{+6}_{-5})~\mathrm{MeV}.
\end{eqnarray}
respectively. 

Recently, the Belle Collaboration reported a new structure named $\Lambda_{c}(2910)$ in the $\Sigma_{c}(2455)^{0,++}\pi^{+,-}$ invariant mass spectrum of the processes $\bar{B}^{0}\to\Sigma_{c}(2455)^{0,++}\pi^{+,-}\bar{p}$~\cite{Belle:2022hnm}. The resonance parameters of this state were reported to be
\begin{eqnarray}
m_{\Lambda_{c}(2910)^{+}}&=&\left(2913.8\pm5.6\pm3.8\right)~\mathrm{MeV},\nonumber\\
\Gamma_{\Lambda_{c}(2910)^{+}}&=&\left(51.8\pm20.0\pm18.8\right)~\mathrm{MeV},
\end{eqnarray}
respectively.

\begin{figure}[t]
  \centering
  \includegraphics[width=8.5cm]{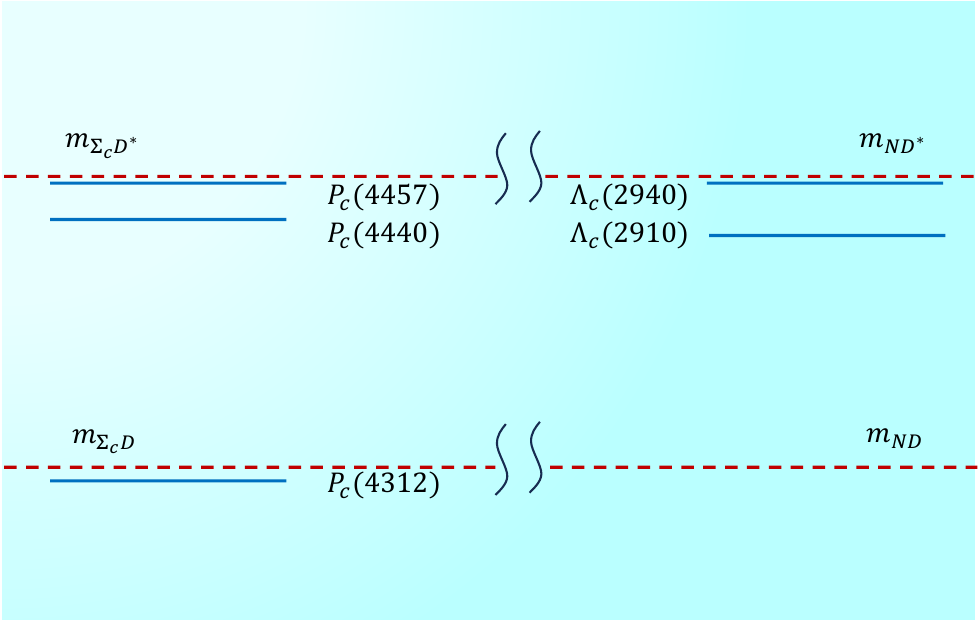}
\caption{A comparison of the $P_c$ states and $\Lambda_c$ states.}\label{Fig:Spectrum}
\end{figure}  

By comparing the masses of $\Lambda_c(2910)$, $\Lambda_c(2940)$ with $P_c$ states, we find some interesting phenomena. As shown in Fig.~\ref{Fig:Spectrum}, the masses of $\Lambda_c(2910)$ and $\Lambda_c(2940)$ are slightly below the threshold of $ND^\ast$, which are similar to the case of $P_c(4440)$ and $P_c(4457)$, respectively. In addition, the masses of $\Lambda_c(2940)$ and $\Lambda_c(2910)$ are about $60\sim 100$ MeV below the quark model expectation for the $2P$ charmed baryon ~\cite{Capstick:1986ter,Ebert:2011kk}. Thus some molecular interpretations have been proposed \cite{Dong:2009tg,Dong:2010xv,Dong:2011ys,Dong:2014ksa,Yan:2022nxp,He:2006is,Zhang:2012jk,Ortega:2012cx,Xin:2023gkf,He:2010zq} besides the conventional charmed baryons~\cite{Zhong:2007gp, Chen:2007xf, Chen:2014nyo,Ebert:2007nw, Azizi:2022dpn, Lu:2018utx,Wang:2022dmw, Zhang:2022pxc}. In particular, the decay properties of $\Lambda_c(2940)$ are discussed in the $ND^{*}$ molecular scenario with different $J^P$ quantum numbers in Ref.~\cite{He:2006is}. The authors in Ref.~\cite{Dong:2009tg} estimated the two-body decays of $\Lambda_{c}(2940)$ through assigning it as $ND^{*}$ molecular with $J^{P}=1/2^{+}$ and $1/2^{-}$ and their results favored $J^{P}=\frac{1}{2}^{+}$. Based on the estimations in Ref.~\cite{Yan:2022nxp}, the authors concluded that $\Lambda_{c}(2940)$ could be interpreted as a $ND^{*}$ molecular state with $J^P=3/2^{-}$, while $\Lambda_{c}(2910)$ cannot be assigned as a molecular state.

Inspired by the recent experimental measurement of $\Lambda_c(2910)$, and the similarity between $\Lambda_c(2910)/\Lambda_c(2940)$ and $P_c(4440)/P_c(4457)$, we consider that both $\Lambda_c(2940)$ and $\Lambda_c(2910)$ are $S$-wave $ND^\ast$ molecular states, and the possible $J^P$ quantum numbers of the molecular states could be $1/2^-$ and $3/2^-$. However, the $J^P$ quantum numbers of $\Lambda_c(2910)$ and $\Lambda_c(2940)$ are not well determined. In the Review of Particle Physics (2023)~\cite{ParticleDataGroup:2022pth}, the $J^P$ quantum numbers of $\Lambda_c(2940)$ are assigned as $3/2^-$, although it is noted that the $J^P=3/2^-$ is favored but not definitive. The determination of $J^P$  quantum numbers for these states is challenging due to the interplay of spin-spin interactions and tensor forces, as discussed in Ref.~\cite{Yamaguchi:2019seo}. Thus, in the present work, we consider two scenarios, A and B, where scenario A corresponds to assuming that the $J^P$ quantum numbers of $\Lambda_c(2910)$ and $\Lambda_c(2940)$ are $1/2^-$ and $3/2^-$, respectively, while scenario B corresponds to the opposite identification. The decays of $\Lambda_c(2910)$ and $\Lambda_c(2940)$ are investigated in the present work, and by comparing with the relevant experiment measurement, one can further check the molecular interpretations and may also provide some helpful information to distinguish two identification scenarios. 

This work is organized as follows. After introduction,  the $S$-wave $ND^\ast $ hadronic molecular structures with different $J^P$ quantum numbers are discussed. The strong decays of $\Lambda_c(2910)$ and $\Lambda_c(2940)$ in the $ND^\ast $ molecular scenarios are estimated in Sec. \ref{Sec:Decay}. The numerical results and the relevant discussions are presented in Sec. \ref{Sec:Results}, and the last section is devoted to a short summary.

\section{Hadronic molecular structure}
\label{Sec:Structure}
In the present work, we take $\Lambda_{c}(2910)$ and $\Lambda_{c}(2940)$ as $N D^{*}$ molecular state, and the possible $J^P$ quantum numbers could be  $1/2^-$ and $3/2^-$. The effective Lagrangian describing the interaction between the $S$-wave molecular states and their components are,
\begin{eqnarray}
\mathcal{L}_{\Lambda_{c}}&=&g_{{\Lambda}_{c}ND^{*}}\bar{\Lambda}_{c}(x)\gamma^{\mu}\gamma^{5}\int dy\Phi(y^2)\left[D^{*0}_{\mu}(x+\omega_{pD^{*0}}y)\right.\nonumber\\&\times&\left.p(x-\omega_{D^{*0}p}y)+D^{*+}_{\mu}(x+\omega_{nD^{*+}}y)n(x-\omega_{D^{*+}n}y)\right]\nonumber\\&+&\mathrm{H.c.},\nonumber\\
\mathcal{L}_{\Lambda_{c}^{\prime}}&=&g_{{\Lambda}_{c}^{\prime}N D^{*}} \bar{\Lambda}_{c}^{\prime}(x)\int dy\Phi(y^2) \left[D^{*0}_{\mu}(x+\omega_{pD^{*0}}y)\right.\nonumber\\&\times&\left.p(x-\omega_{D^{*0}p}y)+D^{*+}_{\mu}(x+\omega_{nD^{*+}}y)n(x-\omega_{D^{*+}n}y)\right]\nonumber\\&+&\mathrm{H.c.},\label{Eq:1}
\end{eqnarray}
where ${\Lambda}_{c}$ and ${\Lambda}_{c}^{\prime}$ indicate the $N D^\ast$ molecular state with $J^P=1/2^-$ and $3/2^-$, respectively, and $\omega_{ij}=\frac{m_{i}}{m_{i}+m_{j}}$. To describe the molecular state interior structure, a correlation function $\Phi(y^{2})$ is introduced, which could also be understood as the wave function of the $D^{*}N$. Considering the similarity of $ND^\ast$ molecular states with different $J^P$ quantum numbers, we use the same correlation function for both $ND^\ast$ molecules with $J^P=1/2^-$ and $3/2^-$. The Fourier transformation of the correlation function is,
\begin{eqnarray}
\Phi(y^{2})=\int \frac{d^{4}p}{(2\pi)^{4}}e^{-ipy}\tilde{\Phi}(-p^{2}).\label{Eq:2}
\end{eqnarray}
The concrete form of the $\tilde{\Phi}(-p^{2})$ should fulfill both conditions that describe the molecular state inner structure and drops fast enough in the ultraviolet region. Here, we use the correlation function of the momentum space in the Gaussian form\cite{Faessler:2007us,Faessler:2007gv,Xiao:2019mvs,Xiao:2016hoa,Yue:2022mnf},
\begin{eqnarray}
\tilde{\Phi}(p_{E}^{2})=\mathrm{exp}\left(-p_{E}^{2}/\Lambda_M^{2}\right),
\label{Eq:Phi}
\end{eqnarray}
where $\Lambda_M$ is a model parameter for describing the distribution of the components in the molecular state, and $P_{E}$ is the Jacobi momentum employed to depict the relative movement between the components of the molecular. 

In the $ND^\ast $ molecular scenario, the coupling constants $g_{{\Lambda}_{c}D^{*}N}$ and $g_{{\Lambda}_{c}^\prime D^{*}N}$ can be estimated by the compositeness condition\cite{vanKolck:2022lqz,Weinberg:1962hj,Salam:1962ap}, which is
\begin{eqnarray}
Z=1-\Pi^{\prime}(m_{\Lambda_{c}^{(\prime)}})=0,
\label{Eq:3}
\end{eqnarray}
where $\Pi^{\prime}(m_{\Lambda_{c}})$ is the derivative of the mass operator of the $\Lambda_{c}$ with $J^{P}=1/2^{-}$. For the state with $J^{P}=3/2^{-}$, the mass operator can be decomposed into the transverse part $\Pi(m_{\Lambda_{c}^{\prime}})$ and the longitudinal part $\Pi^{L}(m_{\Lambda_{c}^{\prime}})$, which is
\begin{eqnarray}
\Pi^{\mu\nu}(m_{\Lambda_{c}}^{\prime})=g^{\mu\nu}_{\bot}\Pi(m_{\Lambda_{c}}^{\prime})+\frac{p^{\mu}p^{\nu}}{p^{2}}\Pi^{L}(m_{\Lambda_{c}}^{\prime}),
\end{eqnarray}
with $g^{\mu\nu}_{\bot}=g^{\mu\nu}-p^{\mu}p^{\nu}/p^{2}$.

\begin{figure}[t]
  \centering
  \includegraphics[width=7cm]{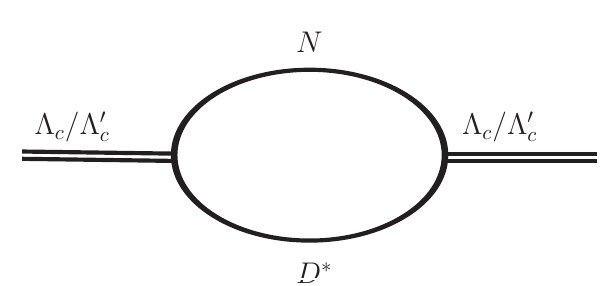}
\caption{Mass operators of $\Lambda_{c}$ and $\Lambda_{c}^{\prime}$ in the $ND^{*}$ molecular frame, with the $J^P$ quantum numbers of $\Lambda_{c}$ and $\Lambda_{c}^{\prime}$ to be $1/2^{-}$ and $3/2^{-}$, respectively.}\label{Fig:Tri1}
\end{figure}

With the effective Lagrangians in Eq.~(\ref{Eq:1}), we can obtain the concrete forms of the mass operator of $\Lambda_{c}$ and $\Lambda_c^\prime$ corresponding to Fig.~\ref{Fig:Tri1}, which read
\begin{eqnarray}
\Pi(m_{\Lambda_{c}})&=&\left(g_{{\Lambda}_{c}ND^{*}}\right)^{2}\int \frac{d^{4}q}{(2\pi)^{4}}\left[\tilde{\Phi}_{\Lambda_{c}}^{2}\left[-(q-\omega_{D^{*0}p}p)^{2},\Lambda_{M}^{2}\right]\right.\nonumber\\&\times&\gamma^{\mu}\gamma^{5}\frac{1}{\slashed{p}-\slashed{q}-m_{p}}\gamma^{\nu}\gamma^{5}\frac{-g^{\mu\nu}+q^{\mu}q^{\nu}/m_{D^{*0}}^{2}}{q^{2}-m_{D^{*0}}^{2}}\nonumber\\&+&\tilde{\Phi}_{\Lambda_{c}}^{2}\left[-(q-\omega_{D^{*+}n}p)^{2},\Lambda_{M}^{2}\right]\nonumber\\&\times&\left.\gamma^{\mu}\gamma^{5}\frac{1}{\slashed{p}-\slashed{q}-m_{n}}\gamma^{\nu}\gamma^{5}\frac{-g^{\mu\nu}+q^{\mu}q^{\nu}/m_{D^{*+}}^{2}}{q^{2}-m_{D^{*+}}^{2}}\right],\nonumber\\
\Pi^{\mu\nu}(m_{\Lambda_{c}^{\prime}})&=&\left(g_{{\Lambda}_{c}^{\prime}ND^{*}}\right)^{2}\int \frac{d^{4}q}{(2\pi)^{4}}\left[\tilde{\Phi}_{\Lambda_{c}^{\prime}}^{2}\left[-(q-\omega_{D^{*0}p}p)^{2},\Lambda_{M}^{2}\right]\right.\nonumber\\&\times&\frac{1}{\slashed{p}-\slashed{q}-m_{p}}\frac{-g^{\mu\nu}+q^{\mu}q^{\nu}/m_{D^{*0}}^{2}}{q^{2}-m_{D^{*0}}^{2}}\nonumber\\&+&\tilde{\Phi}_{\Lambda_{c}^{\prime}}^{2}\left[-(q-\omega_{D^{*+}n}p)^{2},\Lambda_{M}^{2}\right]\nonumber\\&\times&\left.\frac{1}{\slashed{p}-\slashed{q}-m_{n}}\frac{-g^{\mu\nu}+q^{\mu}q^{\nu}/m_{D^{*+}}^{2}}{q^{2}-m_{D^{*+}}^{2}}\right].
\label{Eq:MO}
\end{eqnarray} 

\begin{figure}[t]
\begin{tabular}{cc}
  \centering
 \includegraphics[width=4cm]{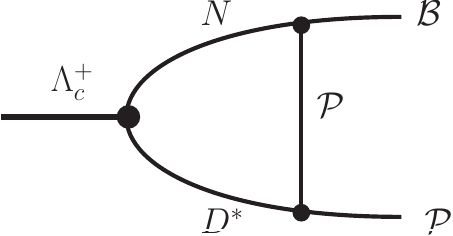}&
 \includegraphics[width=4cm]{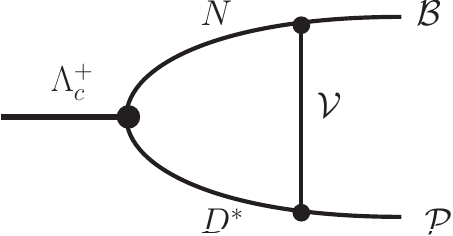}\\
\\
 $(a)$ & $(b)$  \\ \\
  \includegraphics[width=4cm]{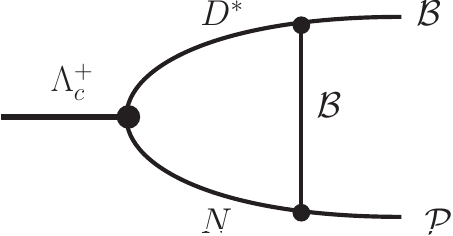}&
 \includegraphics[width=4cm]{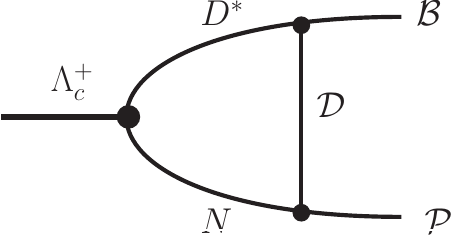}\\
\\ 
 $(c)$ & $(d)$ \\\\
 \includegraphics[width=4cm]{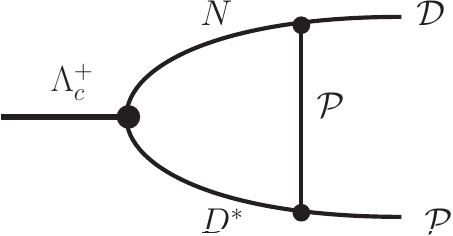}&
\includegraphics[width=4cm]{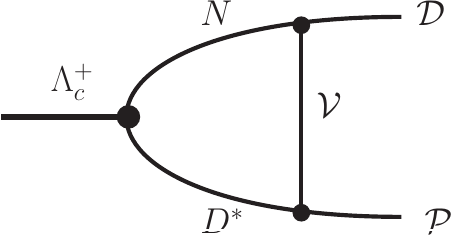}\\
\\
$(e)$&$(f)$\\\\
 \includegraphics[width=4cm]{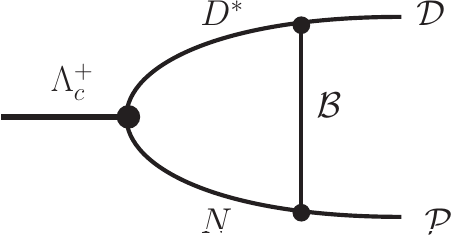}&
\includegraphics[width=4cm]{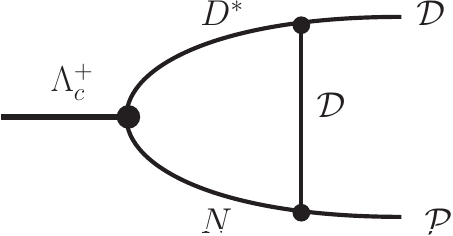}\\
\\
$(g)$&$(h)$\\\\
\end{tabular}

\caption{The typical diagrams contributing to $\Lambda_{c}^{+}$ strong decay at the hadron level.}
\label{Fig:Tri2}
\end{figure}

\section{Strong decays}
\label{Sec:Decay}
In the present work, both $\Lambda_{c}(2940)$ and $\Lambda_{c}(2910)$ are assigned as $S$-wave $ND^{*}$ bound states, and the possible $J^P$ quantum numbers could be $1/2^{-}$ and $3/2^{-}$. With both $J^P$ quantum numbers the $ND^\ast $ molecular state can decay into $ND$ (including $pD^0$ and $n D^+$), $\Sigma_c \pi$ (including $\Sigma_c^0 \pi^+$, $\Sigma_c^+ \pi^0$, and $\Sigma_c^{++}\pi^-$ ), and $\Sigma_c^{\ast} \pi$(including $\Sigma_c^{\ast 0} \pi^+$, $\Sigma_c^{\ast+} \pi^0$, and $\Sigma_c^{\ast ++}\pi^-$). All these decay processes occur in the way that the components, i.e., $ND^\ast$, connect the molecular and the final states by exchanging a proper meson or baryon. All the relevant schematic diagrams contributing to the discussed decay processes are listed in Fig. \ref{Fig:Tri2}, where $\mathcal{P}$ and $\mathcal{V}$ refer to pseudoscalar, and vector meson 15-plets, respectively, while $\mathcal{B}$ and $\mathcal{D}$ denote baryon 20-plets with $J^{P}=1/2^{+}$ and $3/2^{+}$, respectively.

\begin{table*}
\caption{All the possible specific loop contributions to the considered decay channel, where $[ABC]$ indicate the loop that $\Lambda_c(p_0)$ couples to $A(p_1) B(p_2)$, and $A(p_1) B(p_2)$ transits into the final states by exchanging $C(q)$.\label{Tab:Loop}}
 \renewcommand\arraystretch{1.75}
\begin{tabular}{p{1.2cm}<\centering p{2.0cm}<\centering p{11.0cm}<\centering   }
\toprule[1pt]
$J^{P}$ & Final states & Loops \\
\midrule[1pt]
\multirow{4}{*}{$\frac{1}{2}^{-}$}&$ND$&$[ND^{*}\pi]$,$~[ND^{*}\eta]$,$~[ND^{*}\rho]$,$~[ND^{*}\omega]$\\ 

~&$\Sigma_{c}^{(\ast)+}\pi^{0}$&$[pD^{*0}\bar{D}^{0}]$,$~[pD^{*0}\bar{D}^{*0}]$,$~[nD^{*+}D^{-}]$,$~[nD^{*+}D^{*-}]$,$[D^{*0}p\bar{\Delta}^{+}]$,$~[D^{*0}p\bar{p}]$,$~[D^{*+}n\bar{\Delta}^{0}]$,$~[D^{*+}n\bar{n}]$\\
~&$\Sigma_{c}^{(\ast)++}\pi^{-}$&$~[pD^{*0}D^{-}]$,$~[pD^{*0}D^{*-}]$,$[D^{*+}n\bar{\Delta}^{+}]$,$~[D^{*+}n\bar{p}]$,$[D^{*0}p\bar{\Delta}^{++}]$\\
~&$\Sigma_{c}^{(\ast)0}\pi^{+}$&$~[nD^{*+}\bar{D}^{0}]$,$~[nD^{*+}\bar{D}^{*0}]$,$[D^{*0}p\bar{\Delta}^{0}]$,$~[D^{*0}p\bar{n}]$,$[D^{*+}n\bar{\Delta}^{-}]$\\
\midrule[1pt]
\multirow{4}{*}{$\frac{3}{2}^{-}$}&$ND$&$[ND\pi]$,$~[ND\eta]$,$~[ND\rho]$,$~[ND\omega]$\\ 
~&$\Sigma_{c}^{(*)+}\pi^{0}$&$[pD^{*0}\bar{D}^{0}]$,$~[pD^{*0}\bar{D}^{*0}]$,$~[nD^{*+}D^{-}]$,$~[nD^{*+}D^{*-}]$,$[D^{*0}p\bar{\Delta}^{+}]$,$~[D^{*0}p\bar{p}]$,$~[D^{*+}n\bar{\Delta}^{0}]$,$~[D^{*+}n\bar{n}]$\\
~&$\Sigma_{c}^{(*)++}\pi^{-}$&$~[pD^{*0}D^{-}]$,$~[pD^{*0}D^{*-}]$,$[D^{*+}n\bar{\Delta}^{+}]$,$~[D^{*+}n\bar{p}]$,$[D^{*0}p\bar{\Delta}^{++}]$\\
~&$\Sigma_{c}^{(*)0}\pi^{+}$&$~[nD^{*+}\bar{D}^{0}]$,$~[nD^{*+}\bar{D}^{*0}]$,$[D^{*0}p\bar{\Delta}^{0}]$,$~[D^{*0}p\bar{n}]$,$[D^{*+}n\bar{\Delta}^{-}]$\\
\bottomrule[1pt]
\end{tabular}
\end{table*}

\subsection{Effective Lagrangian}

In the present estimations, we employ the effective Lagrangian approach to evaluate the widths of the processes mentioned above. The interaction between molecular states and their components have been presented in Eq.~(\ref{Eq:1}). With SU(4) symmetry, all the other relevant effective Lagrangians read~\cite{Liu:2001yx,Matsuyama:2006rp,Liu:2004mbf,Zhu:2022wpi,Ronchen:2012eg,Hemmert:1997ye,Okubo:1975sc,Shen:2019evi,Lin:2019qiv}
\begin{eqnarray}
\mathcal{L}_{\mathcal{BBP}}&=&-\frac{g_\mathcal{BBP}}{m_{\mathcal{P}}}\bar{\mathcal{B}}\gamma^{\mu}\gamma_{5}\partial_{\mu}\mathcal{PB},\nonumber\\
\mathcal{L}_{\mathcal{BBV}}&=&-g_{\mathcal{BBV}}\bar{\mathcal{B}}\gamma_{\mu}\mathcal{V}^{\mu}\mathcal{B},\nonumber\\
\mathcal{L}_{\mathcal{BDP}}&=&\frac{g_{\mathcal{BDP}}}{m_{\mathcal{P}}}(\bar{\mathcal{D}}^{\mu}\mathcal{B}-\bar{\mathcal{B}}\mathcal{D}^{\mu})\partial_{\mu}\mathcal{P},\nonumber\\
\mathcal{L}_{\mathcal{BDV}}&=&-i\frac{g_{\mathcal{BDV}}}{m_{\mathcal{V}}}(\bar{\mathcal{D}}^{\mu}\gamma^{5}\gamma^{\nu}\mathcal{B}-\bar{\mathcal{B}}\gamma^{5}\gamma^{\nu}\mathcal{D}^{\mu})(\partial_{\mu}\mathcal{V}_{\nu}-\partial_{\nu}\mathcal{V}_{\mu}),\nonumber\\
\mathcal{L}_{\mathcal{DDV}}&=&g_{\mathcal{DDV}}\bar{\mathcal{D}}_{\alpha}\gamma^{\nu}\mathcal{V}_{\nu}\mathcal{D}^{\alpha},\nonumber\\
\mathcal{L}_{\mathcal{PPV}}&=&-ig_{\mathcal{PPV}}(\mathcal{P}\partial_{\mu}\mathcal{P}-\partial_{\mu}\mathcal{PP})\mathcal{V}^{\mu},\nonumber\\
\mathcal{L}_{\mathcal{VVP}}&=&g_{\mathcal{VVP}}\epsilon_{\mu\nu\alpha\beta}\partial^{\mu}\mathcal{V}^{\nu}\partial^{\alpha}\mathcal{V}^{\beta}\mathcal{P}.
\label{Eq:Lag}
\end{eqnarray}
The concrete forms of $\mathcal{P}$, $\mathcal{V}$, $\mathcal{B}$, and $\mathcal{D}$ are listed in Appendix \ref{Sec:App-A}. Note that the relationship between $g_{\mathcal{BBP}}$ and the familiar $g_{A}$ in the $NN\pi$ coupling is given by $g_A = 2 f g_{\mathcal{BBP}} /m_\pi$~\cite{Liu:2011xc}. In the above effective Lagrangian, $m_{\mathcal{P}}$ and $m_{\mathcal{V}}$ refer to the masses of the corresponding pseudoscalar and vector mesons. The $\mathrm{SU}(4)$ symmetry is explicitly broken by the involved hadron masses, for instance, the mass of the $\rho$, $K^\ast$ and $D^\ast$ should be used in the relevant effective Lagrangian.

\subsection{Decay amplitude}
According to the effective Lagrangians listed above, we can obtain the amplitudes for  $\Lambda_c$ decays corresponding to the diagrams in Fig.~\ref{Fig:Tri2}, which are
\begin{eqnarray}
i\mathcal{M}^{(a)}_{[ND^{*}\mathcal{P}]}&=&i^{3}\int\frac{d^{4}q}{(2\pi^{4})}\tilde{\Phi}(-p_{12}^{2},\Lambda_{M}^{2})g_{{\Lambda}_{c}ND^{*}}g_{\mathcal{BBP}}g_{\mathcal{VPP}}\nonumber\\&\times&(i)^3\bar{u}(p_{3})\gamma^{\mu}\gamma^{5}(iq)_{\mu}\frac{1}{\slashed{p}_{1}-m_{1}}\gamma^{\nu}\gamma^{5}g_{\nu\tau}u(p_{0})\nonumber\\&\times&(ip_{4}+iq)^{\alpha}g_{\alpha \beta}\frac{-g^{\tau \beta}+p_{2}^{\tau}p_{2}^{\beta}/m_{2}^{2}}{p_{2}^{2}-m_{2}^{2}}\frac{1}{q^{2}-m_{2}^{2}}\nonumber\\&\times&\mathcal{F}(m_{q},\Lambda),\nonumber
\end{eqnarray}
\begin{eqnarray}
i\mathcal{M}^{(b)}_{[ND^{*}\mathcal{V}]}&=&i^{3}\int\frac{d^{4}q}{(2\pi^{4})}\tilde{\Phi}(-p_{12}^{2},\Lambda_{M}^{2})g_{{\Lambda}_{c}ND^{*}}g_{\mathcal{BBV}}g_{\mathcal{VVP}}\nonumber\\&\times&(i)^3\bar{u}(p_{3})\gamma^{\lambda}\frac{1}{\slashed{p}_{1}-m_{1}}\gamma_{\nu}\gamma^{5}g^{\nu\tau}u(p_{0})\epsilon_{\alpha\beta\rho\theta}(-ip_{2}^{\alpha})\nonumber\\&\times&(-iq^{\rho})g^{\beta \phi}g^{\theta\xi}\frac{-g_{\tau \phi}+p_{2\tau}p_{2\phi}/m_{2}^{2}}{p_{2}^{2}-m_{2}^{2}}\frac{-g_{\lambda\xi}+q_{\lambda}q_{\xi}/m_{2}^{2}}{q^{2}-m_{q}^{2}}\nonumber\\&\times&\mathcal{F}(m_{q},\Lambda),\nonumber
\end{eqnarray}
\begin{eqnarray}
i\mathcal{M}^{(c)}_{[D^{*}N\mathcal{B}]}&=&i^{3}\int\frac{d^{4}q}{(2\pi^{4})}\tilde{\Phi}(-p_{12}^{2},\Lambda^{2}_{M})g_{{\Lambda}_{c}ND^{*}}g_{\mathcal{BBV}}g_{\mathcal{BBP}}\nonumber\\&\times&(i)^3\bar{u}(p_{3})\gamma_{\delta}\frac{1}{\slashed{q}-m_{q}}\gamma^{\mu}\gamma^{5}(ip_{4\mu})\frac{1}{\slashed{p}_{2}-m_{2}}\gamma^{\nu}\gamma^{5}\nonumber\\&\times&g_{\nu\phi}u(p_{0})\frac{-g^{\phi\delta}+p_{1}^{\phi}p_{1}^{\delta}/m_{1}^{2}}{p_{1}^{2}-m_{1}^{2}}\mathcal{F}(m_{q},\Lambda),\nonumber
\end{eqnarray}
\begin{eqnarray}
i\mathcal{M}^{(d)}_{[D^{*}N\mathcal{D}]}&=&i^{3}\int\frac{d^{4}q}{(2\pi^{4})}\tilde{\Phi}(-p_{12}^{2},\Lambda^{2}_{M})g_{{\Lambda}_{c}ND^{*}}g_{\mathcal{BDV}}g_{\mathcal{BDP}}\nonumber\\&\times&(i)^3\bar{u}(p_{3})g^{\mu\lambda}\gamma^{5}\gamma^{\nu}((-ip_{1\mu})g_{\nu\delta}-(-ip_{1\nu})g_{\mu\delta})\nonumber\\&\times&\frac{G_{\lambda\xi}(q,m_q)}{\slashed{q}-m_{q}}ip_{4}^{\xi}\frac{1}{\slashed{p}_2-m_2}\gamma^{\alpha}\gamma^{5}g_{\alpha\phi}u(p_{0})\nonumber\\&\times&\frac{-g^{\phi\delta}+p_{1}^{\phi}p_{1}^{\delta}/m_{1}^{2}}{p_{1}^{2}-m_{1}^{2}}\mathcal{F}(m_{q},\Lambda),\nonumber
\end{eqnarray}
\begin{eqnarray}
i\mathcal{M}^{(e)}_{[ND^{*}\mathcal{P}]}&=&i^{3}\int\frac{d^{4}q}{(2\pi^{4})}\tilde{\Phi}(-p_{12}^{2},\Lambda_{M}^{2})g_{{\Lambda}_{c}ND^{*}}g_{\mathcal{BDP}}g_{\mathcal{VPP}}\nonumber\\&\times&(i)^{3}\bar{u}^{\theta}(p_{3})g_{\mu\theta}iq^{\mu}\frac{1}{\slashed{p}_{1}-m_{1}}\gamma^{\nu}\gamma^{5}g_{\nu \tau}u(p_{0})\nonumber\\&\times&(ip_{4}^{\alpha}+iq^{\alpha})g_{\alpha \phi}\frac{-g^{\tau \phi}+p_{2}^{\tau}p_{2}^{\phi}/m_{2}^{2}}{p_{2}^{2}-m_{2}^{2}}\frac{1}{q^{2}-m_{q}^{2}}\nonumber\\&\times&\mathcal{F}(m_{q},\Lambda),\nonumber
\end{eqnarray}
\begin{eqnarray}
i\mathcal{M}^{(f)}_{[ND^{*}\mathcal{V}]}&=&i^{3}\int\frac{d^{4}q}{(2\pi^{4})}\tilde{\Phi}(-p_{12}^{2},\Lambda_{M}^{2})g_{{\Lambda}_{c}ND^{*}}g_{\mathcal{BDV}}g_{\mathcal{VVP}}\nonumber\\&\times&(i)^3\bar{u}^{\theta}(p_{3})\gamma^{5}\gamma_{\nu}g_{\mu\theta}(iq^{\mu}g^{\nu\lambda}-iq^{\nu}g^{\mu\lambda})\frac{1}{\slashed{p}_{1}-m_{1}}\nonumber\\&\times&\gamma_{\alpha}\gamma^{5}g^{\alpha \tau}u(p_{0})\epsilon_{\beta\rho\omega\sigma}(-ip_{2}^{\beta})(-iq)^{\omega}g^{\rho \phi}g^{\sigma\xi}\nonumber\\&\times&\frac{-g_{\tau \phi}+p_{2\tau}p_{2\phi}/m_{2}^{2}}{p_{2}^{2}-m_{2}^{2}}\frac{-g_{\lambda\xi}+q_{\lambda}q_{\xi}/m_{q}^{2}}{q^{2}-m_{q}^{2}}\mathcal{F}(m_{q},\Lambda),\nonumber
\end{eqnarray}
\begin{eqnarray}
i\mathcal{M}^{(g)}_{[D^{*}N\mathcal{B}]}&=&i^{3}\int\frac{d^{4}q}{(2\pi^{4})}\tilde{\Phi}(-p_{12}^{2},\Lambda_{M}^{2})g_{{\Lambda}_{c}ND^{*}}g_{\mathcal{BDV}}g_{\mathcal{BBP}}\nonumber\\&\times&(i)^3\bar{u}_{\theta}(p_{3})\gamma^{5}\gamma^{\nu}g^{\mu\theta}(-ip_{1\mu}g_{\nu\delta}+ip_{1\nu}g_{\mu\delta})\nonumber\\&\times&\frac{1}{\slashed{q}-m_{q}}\gamma_{\alpha}\gamma^{5}(ip_{4}^{\alpha})\frac{1}{\slashed{p}_{2}-m_{2}}\gamma^{\beta}\gamma^{5}g_{\beta\phi}u(p_{0})\nonumber\\&\times&\frac{-g^{\phi\delta}+p_{1}^{\phi}p_{1}^{\delta}/m_{1}^{2}}{p_{1}^{2}-m_{1}^{2}}\mathcal{F}(m_{q},\Lambda),\nonumber
\end{eqnarray}
\begin{eqnarray}
i\mathcal{M}^{(h)}_{[D^{*}N\mathcal{D}]}&=&i^{3}\int\frac{d^{4}q}{(2\pi^{4})}\tilde{\Phi}(-p_{12}^{2},\Lambda_{M}^{2})g_{{\Lambda}_{c}ND^{*}}g_{\mathcal{DDV}}g_{\mathcal{BDP}}\nonumber\\&\times&(i)^3\bar{u}_{\theta}(p_{3})g_{\alpha\lambda}g^{\alpha\theta}\gamma_{\delta}\frac{G^{\lambda\xi}(q,m_{q})}{\slashed{q}-m_{q}}(ip_{4\xi})\nonumber\\&\times&\frac{1}{\slashed{p}_{2}-m_{2}}\gamma^{\nu}\gamma^{5}g_{\nu\phi}u({p_{0}})\frac{-g^{\phi\delta}+p_{1}^{\phi}p_{1}^{\delta}/m_{1}^{2}}{p_{1}^{2}-m_{1}^{2}}\nonumber\\&\times&\mathcal{F}(m_{q},\Lambda),
\end{eqnarray}
where the $G_{\mu \nu}(p, m)$ is the numerator of the propagator for the baryon with spin $3/2$,  which is
\begin{eqnarray}
 G_{\mu \nu}(p, m)=-g_{\mu v}+\frac{1}{3} \gamma_{\mu} \gamma_{\nu}+\frac{1}{3 m}\left(\gamma_{\mu} p_{\nu}-\gamma_{\nu} p_{\mu}\right)+\frac{2}{3} \frac{p_{\mu} p_{v}}{m^{2}}.\nonumber 
\end{eqnarray}

In the above amplitudes, a form factor is introduced to describe the exchanging meson inner structure and off-shell effect, which is
\begin{eqnarray}
\mathcal{F}\left(m_{q},\Lambda\right)=\frac{\Lambda^{4}}{(m^{2}-q^{2})^{2}+\Lambda^{4}},\label{Eq:FFs}
\end{eqnarray}
with $\Lambda$ to be a model parameter, which should be of the order of 1 GeV.

For the specific decay process, we list all the possible loop contributions in Table~\ref{Tab:Loop}, and the total amplitudes for each channel are
\begin{eqnarray}
\mathcal{M}_{\Lambda_{c}\to ND}^{tot}&=&\mathcal{M}_{[pD^{*0}\pi]}^{(a)}+\mathcal{M}_{[pD^{*0}\eta]}^{(a)}+\mathcal{M}_{[pD^{*0}\omega]}^{(b)}+\mathcal{M}_{[pD^{*0}\rho]}^{(b)},\nonumber\\
\mathcal{M}_{\Lambda_{c}\to \Sigma_{c}^{+}\pi^{0}}^{tot}&=&\mathcal{M}_{[pD^{*0}\bar{D}^{0}]}^{(a)}+\mathcal{M}_{[pD^{*0}\bar{D}^{*0}]}^{(b)}+\mathcal{M}_{[nD^{*+}D^{-}]}^{(a)}\nonumber
\\&+&\mathcal{M}_{[nD^{*+}D^{*-}]}^{(b)}+\mathcal{M}_{[D^{*0}p\bar{\Delta}^{+}]}^{(d)}+\mathcal{M}_{[D^{*0}p\bar{p}]}^{(c)}\nonumber\\&+&\mathcal{M}_{[D^{*+}n\bar{\Delta}^{0}]}^{(d)}+\mathcal{M}_{[D^{*+}n\bar{n}]}^{(c)},\nonumber\\
\mathcal{M}_{\Lambda_{c}\to \Sigma_{c}^{++}\pi^{-}}^{tot}&=&\mathcal{M}_{[pD^{*0}\bar{D}^{-}]}^{(a)}+\mathcal{M}_{[pD^{*0}\bar{D}^{*-}]}^{(b)}+\mathcal{M}_{[D^{*+}n\bar{\Delta}^{+}]}^{(d)}\nonumber\\&+& \mathcal{M}_{[D^{*+}n\bar{p}]}^{(c)}+\mathcal{M}_{[D^{*0}p\bar{\Delta}^{++}]}^{(d)},\nonumber\\
\mathcal{M}_{\Lambda_{c}\to \Sigma_{c}^{*0}\pi^{+}}^{tot}&=&\mathcal{M}_{[D^{*0}p\bar{\Delta}^{0}]}^{(d)}+\mathcal{M}_{[D^{*0}p\bar{n}]}^{(c)}+\mathcal{M}_{[nD^{*+}\bar{D}^{0}]}^{(a)}\nonumber\\&+&\mathcal{M}_{[nD^{*+}\bar{D}^{*0}]}^{(b)}+\mathcal{M}_{[D^{*+}n\bar{\Delta}^{-}]}^{(d)},\nonumber\\
\mathcal{M}_{\Lambda_{c}\to \Sigma_{c}^{*+}\pi^{0}}^{tot}&=&\mathcal{M}_{[pD^{*0}\bar{D}^{0}]}^{(e)}+\mathcal{M}_{[pD^{*0}\bar{D}^{*0}]}^{(f)}+\mathcal{M}_{[nD^{*+}D^{-}]}^{(e)}\nonumber\\&+&\mathcal{M}_{[nD^{*+}D^{*-}]}^{(f)}+\mathcal{M}_{[D^{*0}p\bar{\Delta}^{+}]}^{(h)}+\mathcal{M}_{[D^{*0}p\bar{p}]}^{(g)}\nonumber\\&+&\mathcal{M}_{[D^{*+}n\bar{\Delta}^{0}]}^{(h)}+\mathcal{M}_{[D^{*+}n\bar{n}]}^{(g)}\nonumber,\\
\mathcal{M}_{\Lambda_{c}\to \Sigma_{c}^{*++}\pi^{-}}^{tot}&=&\mathcal{M}_{[pD^{*0}\bar{D}^{-}]}^{(e)}+\mathcal{M}_{[pD^{*0}\bar{D}^{*-}]}^{(f)}+\mathcal{M}_{[D^{*+}n\bar{\Delta}^{+}]}^{(h)}\nonumber\\&+& \mathcal{M}_{[D^{*+}n\bar{p}]}^{(g)}+\mathcal{M}_{[D^{*0}p\bar{\Delta}^{++}]}^{(h)},\nonumber\\
\mathcal{M}_{\Lambda_{c}\to \Sigma_{c}^{0}\pi^{+}}^{tot}&=&\mathcal{M}_{[D^{*0}p\bar{\Delta}^{0}]}^{(h)}+\mathcal{M}_{[D^{*0}p\bar{n}]}^{(g)}+\mathcal{M}_{[nD^{*+}\bar{D}^{0}]}^{(e)}\nonumber\\&+&\mathcal{M}_{[nD^{*+}\bar{D}^{*0}]}^{(f)}+\mathcal{M}_{[D^{*+}n\bar{\Delta}^{-}]}^{(h)}.
\end{eqnarray}
With the total amplitudes listed above, we can estimate the partial widths of the decay processes listed in Table~\ref{Tab:Loop} by
\begin{eqnarray}
\Gamma_{\Lambda_{c}\to ...}&=&\frac{1}{(2J+1)\pi}\frac{|\vec{p}|}{m_{\Lambda_{c}}^{2}}\left|\ \overline{\mathcal{M}_{\Lambda_{c}\to ...}}\ \right|^{2}.
\end{eqnarray}

In the same way, one can obtain the decay amplitudes for the $ND^\ast$ molecular state with $J^{P}=3/2^{-}$ and then estimate the corresponding partial widths with the above equation.

\section{Numerical results and discussions}
\label{Sec:Results}
\subsection{Coupling constants under SU(4) symmetry}
Under SU(4) symmetry, the coupling constants $g_{\mathcal{PPV}}$ and $g_{\mathcal{VVP}}$ could be related by the same gauge coupling constant $g$ by~\cite{Yue:2022mnf,Xiao:2020ltm,Chen:2011cj,Oh:2000qr,Kaymakcalan:1984bz,Lin:1999ad}
\begin{eqnarray}
g_{VPP}=\frac{1}{4}g ,~~~~~~~~g_{VVP}=\frac{1}{4}\frac{g^{2}N_{c}}{16\pi^{2}f_{\pi}},
\end{eqnarray}
where $f_{\pi}=132$ MeV is the decay constant of the $\pi$ meson; $g=12.48$, which can be derived by the observed decay width of the process $K^{*}\to K\pi$~\cite{ParticleDataGroup:2022pth,Yue:2022mnf}.

 Regarding the coupling constants related to baryons, the following values are utilized~\cite{Zhu:2022wpi,Hofmann:2005sw,Kong:2023dwz,Ronchen:2012eg,Hemmert:1997ye,Shen:2019evi,Liu:2001yx,Xiao:2020alj}
\begin{eqnarray}
g_{\mathcal{BBP}}&=&g_{NN\pi}=0.989,\nonumber\\
g_{\mathcal{BBV}}&=&g_{NN\rho}=3.25,\nonumber\\
g_{\mathcal{BDP}}&=&g_{\Delta N\pi}=2.12,\nonumber\\
g_{\mathcal{BDV}}&=&g_{\Delta N\rho}=6.08,\nonumber\\
g_{\mathcal{DDV}}&=&g_{\rho\Delta\Delta}=-4.3.
\end{eqnarray}
Other coupling constants of those type can be derived from $\mathrm{SU}(4)$ symmetry.

\begin{figure}[t]
  \centering
  \includegraphics[width=8.2 cm]{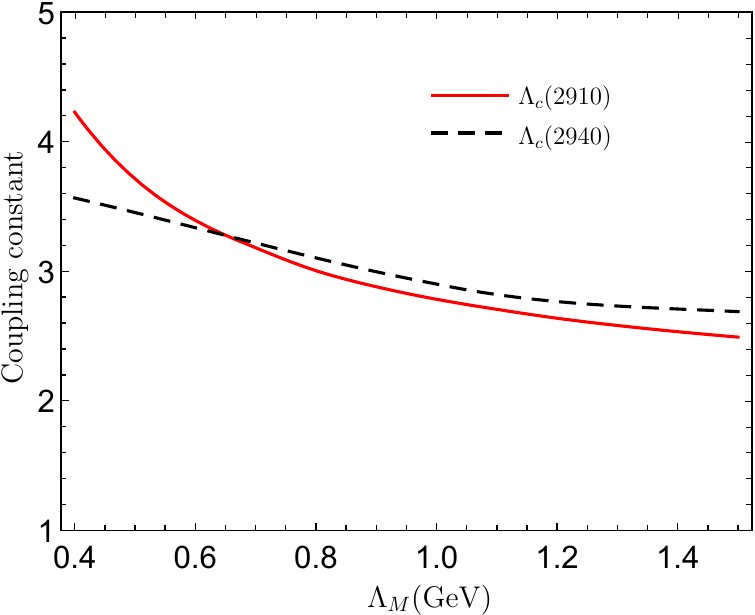}
\caption{The coupling constants $g_{\Lambda_c N D^\ast}$ and $g_{\Lambda_c^\prime ND^\ast}$ depending on the model parameter $\Lambda_M$ in scenario A.\label{Fig:CP-A}}
\end{figure}

\subsection{Scenario A}
In scenario A, we consider $\Lambda_c(2910)$ and $\Lambda_c(2940)$ as $ND^\ast $ molecular states with $J^P=1/2^-$ and $3/2^-$, respectively. The coupling constants of the molecular states and its components could be estimated by the Weinberg's compositeness condition as indicated in Eq.~(\ref{Eq:3}). The coupling constants $g_{\Lambda_c ND^\ast }$ and $g_{\Lambda_c^\prime ND^\ast}$ depending on the model parameter $\Lambda_M$ are presented in Fig.~\ref{Fig:CP-A}, where  $\Lambda_M$ varies from 0.4 to 1.5 GeV. Our estimations indicate that the coupling constants smoothly decrease with the increasing of $\Lambda_M$. As indicated in Eq.~\eqref{Eq:Phi}, the correlation function or, equivalently, the wave function will decrease at a slower rate with larger $\Lambda_M$, suggesting that the molecular components interact over a larger spatial region. From Eq.~\eqref{Eq:MO}, it is evident that the loop integral part of the mass operator increases with higher values of $\Lambda_M$. Consequently, the coupling constants exhibit a decreasing trend as $\Lambda_M$ increases. In the considered $\Lambda_M$ range, the coupling constants $g_{\Lambda_c ND^\ast N}$ and $g_{\Lambda_c^\prime N D^\ast}$ are estimated to be $4.22 \sim 2.49$ and $3.56 \sim 2.69$, respectively.

\begin{figure}[t]
  \centering
  \includegraphics[width=8.5 cm]{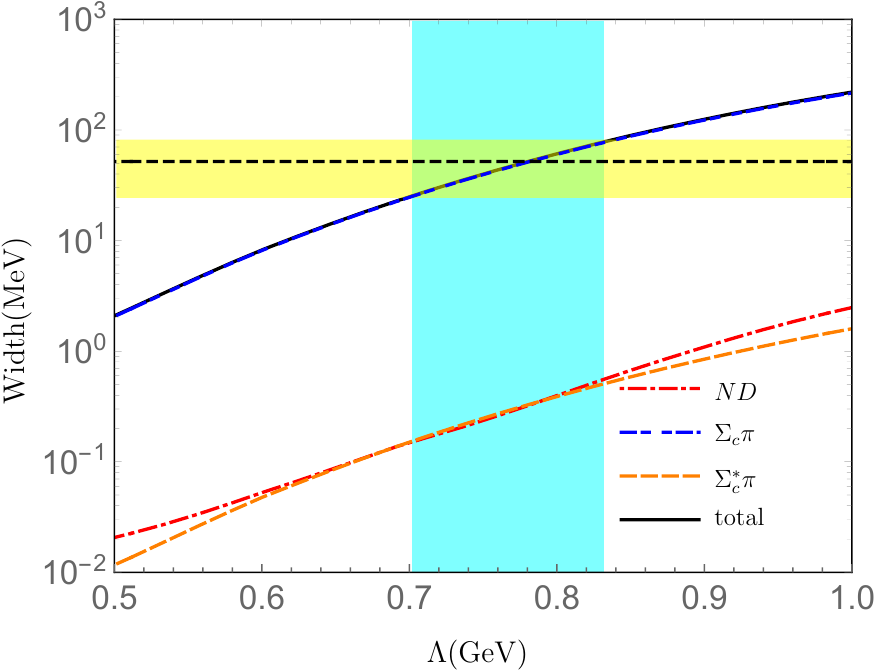}
\caption{The partial widths of $\Lambda_{c}(2910)\to ND$, $\Sigma_{c}\pi$, and $\Sigma_c^\ast \pi$ depending on the parameter $\Lambda$, where $\Lambda_c(2910)$ is assigned as $D^\ast N$ molecular state with $J^{p}=1/2^{-}$. }\label{Fig:L2910-A}
\end{figure}  

\begin{figure}[t]
  \centering
  \includegraphics[width=8.5cm]{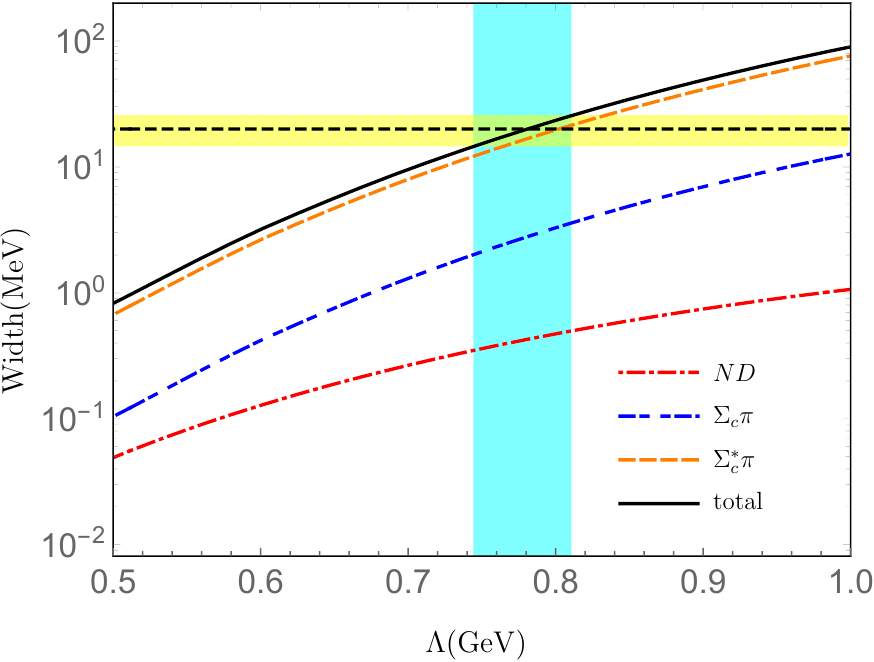}
\caption{The partial widths of $\Lambda_{c}(2940)\to ND$, $\Sigma_{c}\pi$, and $\Sigma_c^\ast \pi$ depending on parameter $\Lambda$, where $\Lambda_c(2940)$ is taken as $D^\ast N$ molecular state with $J^{p}=2/3^{-}$.}\label{Fig:L2940-A}
\end{figure}

\begin{table*}[htb]
\renewcommand\arraystretch{1.75}
\caption{The branching fractions for $ND$, $\Sigma_c \pi$ and $\Sigma_c^\ast \pi$ channels for $\Lambda_c(2910)$ and $\Lambda_c(2940)$ in different scenarios.  \label{Tab:Br}}
\begin{tabular}{p{2cm}<\centering p{3cm}<\centering p{3cm}<\centering p{0.2cm}<\centering p{3cm}<\centering p{3cm}<\centering}
\toprule[1pt]
\multirow{2}*{Channel} & \multicolumn{2}{c}{Scenario A} &&\multicolumn{2}{c}{Scenario B}\\
\cmidrule[1pt]{2-3}  \cmidrule[1pt]{5-6}
    & $\Lambda_c(2910)$ & $\Lambda_c(2940)$ && $\Lambda_c(2940)$ &$\Lambda_c(2910)$ \\
 \midrule[1pt]
 $ND$ &  $(4.01\sim 5.62)\times 10^{-3} $ & $(1.90\sim 2.40)\%$  && $(3.49\sim 3.90)\times10^{-3}  $ & $(1.44 \sim 1.49)\% $\\
 $\Sigma_c \pi$ & $(98.7\sim 98.9)\%$ & $(13.8\sim 14.0)\%$ && $\sim 98.8\%$  &$(12.4\sim 12.5)\%$\\
 $\Sigma_c^\ast \pi $ & $(6.18\sim 6.43)\times10^{-3}$ & $(83.7\sim 84.0)\%$ && $(7.72\sim 7.76)\times10^{-3}$ & $\sim 86.0\%$\\
\bottomrule[1pt]
\end{tabular} 
\end{table*}

With the above preparations, we can estimate the partial widths of the relevant processes. In this work, the $ND$, $\Sigma_c\pi$ and $\Sigma_c^\ast \pi$ channels are taken into consideration, which should be the dominant decay channels of both $\Lambda_c(2940)$ and $\Lambda_c(2910)$. In the present estimations, two model parameters are introduced, which are $\Lambda_M$ introduced by the correlation function and $\Lambda$ employed by the form factor.  Here we take $\Lambda_M=0.7$ GeV and vary $\Lambda$ from $0.5$ to 1.0 GeV. The widths of considered channels depending on the model parameter $\Lambda$ are presented in Figs.~\ref{Fig:L2910-A} and \ref{Fig:L2940-A}, where the widths increase with the increasing of $\Lambda$. The parameter $\Lambda$ solely emerges in the form factor $\mathcal{F}(m_q,\Lambda)$. The explicit expression of $\mathcal{F}(m_q,\Lambda)$ in Eq. \eqref{Eq:FFs} reveals that it amplifies with the increasing of $\Lambda$ . Consequently, the estimated partial decay widths, as well as the total widths, also increase with the rising of $\Lambda$. From the figures one can find that the measured widths of $\Lambda_c(2910)$ and $\Lambda_c(2940)$ can be well reproduced in the range $0.70~<\Lambda<0.83$  and $0.74<\Lambda<0.81~\mathrm{GeV}$, respectively.

In the common parameter range, i.e., $0.74<\Lambda <0.81 ~\mathrm{GeV}$, the estimated branching fractions for the considered channels are collected in Table~\ref{Tab:Br}. Our estimations indicate that the $\Lambda_c(2910)$ dominantly decays into $\Sigma_c \pi$ with a branching fraction $(98.7\sim 98.9)\%$, while the branching fractions for $\Sigma_c^\ast \pi$     and $ND$ are rather small, which are $(6.18\sim 6.43)\times 10^{-3}$ and $(4.01\sim 5.62)\times 10^{-3}$, respectively. In the same parameter range, we find the branching fractions of $\Sigma_c^\ast \pi$, $\Sigma_c \pi$, and $ND$ channels for $\Lambda_{c}(2940)$ are $(83.7\sim 84.0)\%$, $(13.8\sim 14.0)\%$, and $(1.90\sim 2.40)\%$, respectively. From the present estimations, we can conclude that the observations of $\Lambda_c(2910)$ in the $ND$ and $\Sigma_c^\ast \pi$ channels are rather difficult. In the $\Sigma_c \pi$ channel, both $\Lambda_c(2910)$ and $\Lambda_c(2940)$ could be observed, and in the $ND$ channel, $\Lambda_c(2940)$ is more likely to be observed.

\begin{figure}[t]
  \centering
  \includegraphics[width=8.2cm]{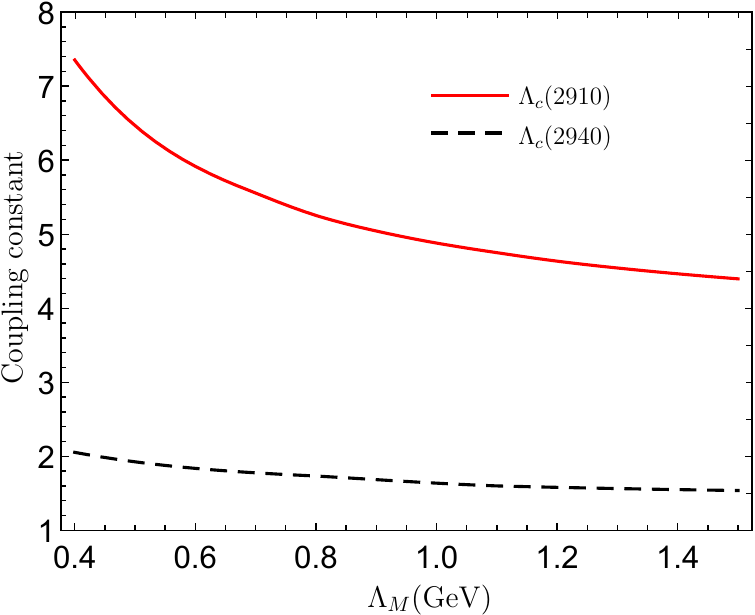}
\caption{The coupling constants $g_{\Lambda_c D^\ast N}^{1/2}$ and $g_{\Lambda_c D^\ast N}^{3/2}$ depending on the model parameter $\Lambda_M$ in scenario B.\label{Fig:CP-B}}
\end{figure}

\subsection{Scenario B}

In scenario B, we consider $\Lambda_c(2940)$ and $\Lambda_c(2910)$ as $ND^\ast$ molecular states with $J^P=1/2^-$ and $3/2^-$, respectively. The coupling constants of the molecular states and its components depending on the model parameter $\Lambda_M$ are presented in Fig.~\ref{Fig:CP-B}. In scenario B, $g_{\Lambda_c^\prime  ND^\ast }$ is several times larger than $g_{\Lambda_c D^\ast N}$. In particular, in the considered parameter range, i.e., $0.4<\Lambda_M<1.5~\mathrm{GeV}$, $g_{\Lambda_c ND^\ast}$ decreases from $2.05$ to $1.54$, while $g_{\Lambda_c^\prime ND^\ast}$ varies from $7.35$ to $4.39$.

\begin{figure}[t]
  \centering
  \includegraphics[width=8.5 cm]{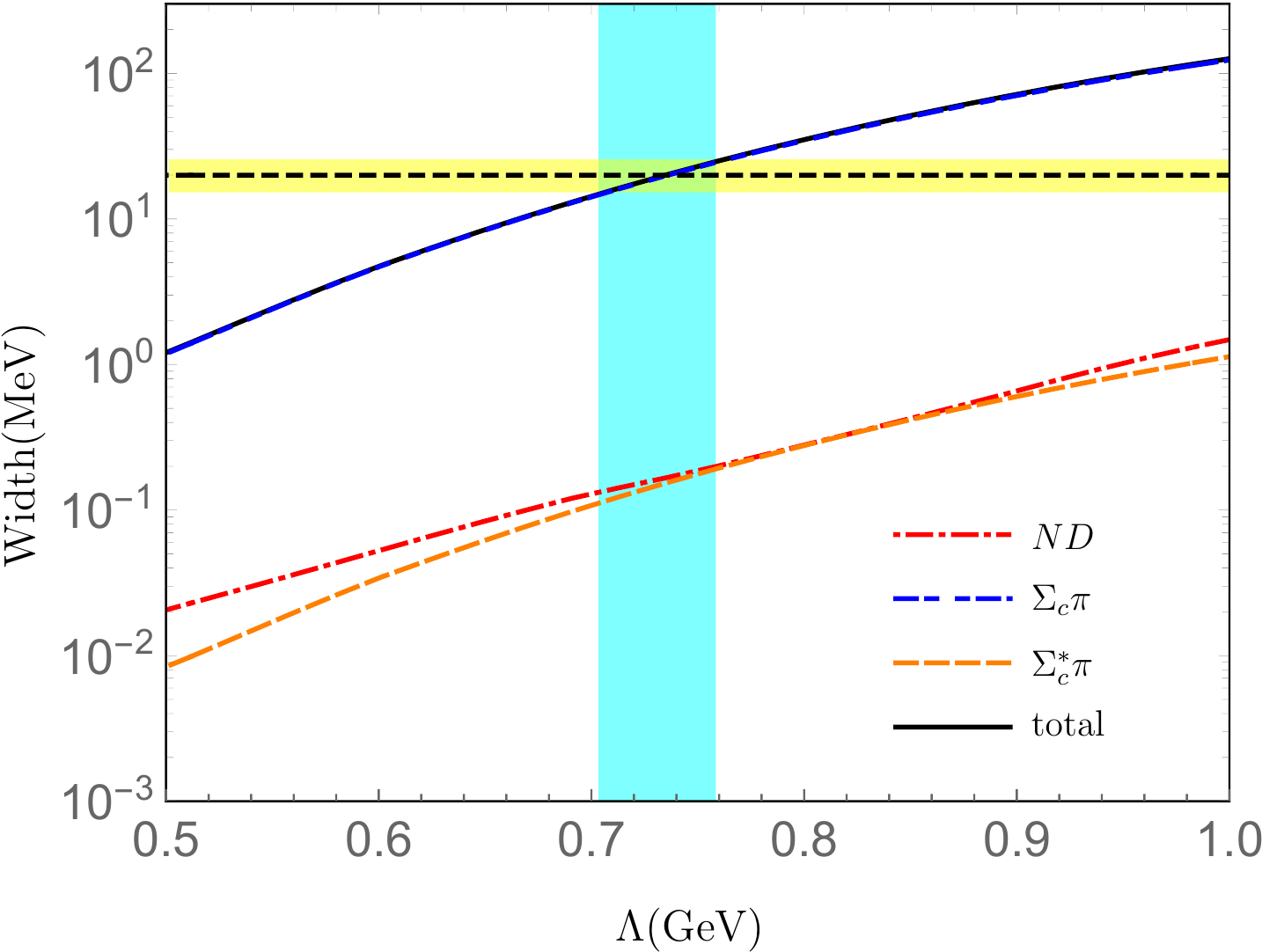}
\caption{The partial widths of $\Lambda_{c}(2940)\to ND$, $\Sigma_{c}\pi$, $\Sigma_{c}^{*}\pi$ depending on the parameter $\Lambda$, where $\Lambda_c(2940)$ is considered as a $D^\ast N$ molecular state with $J^{p}=1/2^{-}$. \label{Fig:L2940-B}}
\end{figure}  

\begin{figure}[t]
  \centering
  \includegraphics[width=8.5 cm]{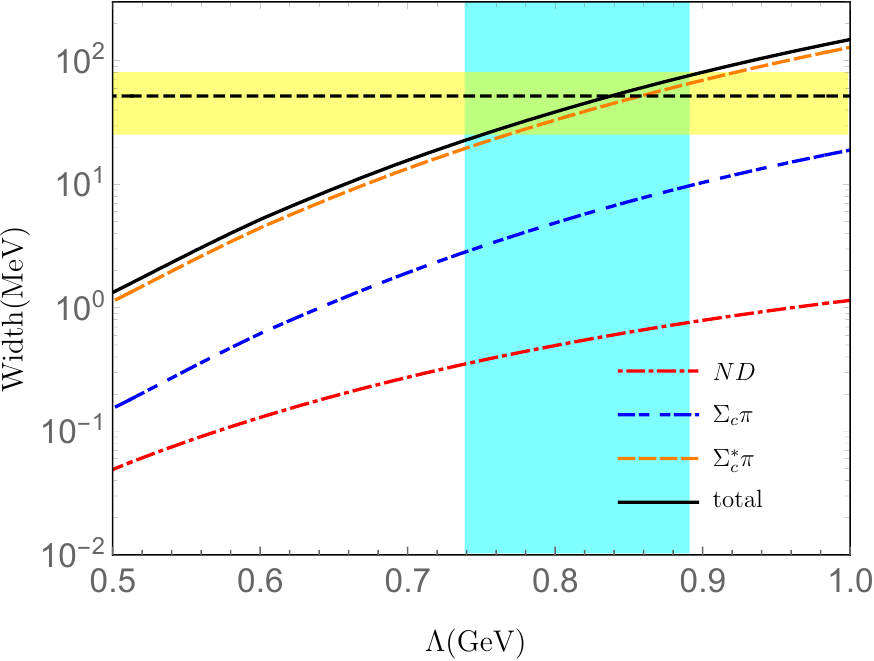}
\caption{The partial widths of $\Lambda_{c}(2910)\to ND$, $\Sigma_{c}\pi$, $\Sigma_{c}^{*}\pi$ depending on the parameter $\Lambda$, where $\Lambda_c(2910)$ is considered as a $D^\ast N$ molecular state with $J^{p}=3/2^{-}$. \label{Fig:L2910-B}.}\end{figure}

The partial widths of the relevant channels for $\Lambda_c(2940)$ and $\Lambda_c(2910)$ depending on the parameter $\Lambda$ are presented in Figs.~\ref{Fig:L2940-B} and ~\ref{Fig:L2910-B}, respectively. Our estimations indicate that in scenario B, the widths of $\Lambda_c(2940)$ and $\Lambda_c(2910)$ can be reproduced in the  ranges of $(0.7<\Lambda< 0.76)$ and $(0.75 <\Lambda< 0.89)$ $\mathrm{GeV}$, respectively. In a very small common parameter range, i.e., $0.75<\Lambda<0.76 ~\mathrm{GeV}$, the branching fractions of the $ND$, $\Sigma_c \pi$, and $\Sigma_c^\ast \pi$ channels are collected in Table~\ref{Tab:Br}. As for $\Lambda_c(2940)$, its dominant channel is $\Sigma_c \pi$ with the branching ratio to be $98.8\%$, and the branching ratios of the $ND$ and $\Sigma_c^\ast \pi $ channels are estimated to be $(3.49\sim 3.90) \times 10^{-3}$ and $(7.72\sim 7.76)\times 10^{-3}$, respectively. In addition, the branching fractions of $\Lambda_c(2910) \to ND$, $\Sigma_c \pi$, and $\Sigma_c^\ast \pi$ are evaluated to be $(1.44\sim 1.49)\%$, $(12.4\sim 12.5)\%$, and $~86.0\%$, respectively.

It should be noted that $\Lambda_c(2940)$ was first observed in the $D^0 p$ invariant mass spectrum; the branching fractions of $\Lambda_c(2940) \to ND$ are estimated to be $(1.90\sim 2.40)\%$ and $(3.49\sim 3.90) \times 10^{-3}$ in scenario A and scenario B, respectively. Thus, from the decay properties of $\Lambda_c(2940)$, scenario A is weakly favored. As for the $\Sigma_c \pi$ channel, our estimations indicate that the branching fractions for $\Lambda_c\to \Sigma_c \pi$ are larger than the one of $\Lambda_c^\prime \to \Sigma_c \pi$, however, the branching ratio of $\Lambda_c^\prime\to \Sigma_c \pi$ is more than $10\%$, which should be also sizable to be observed, thus, the structure in $\Sigma_c \pi$ invariant mass distributions may contain both $\Lambda_c(2940)$ and $\Lambda_c(2910)$. In addition, we find the branching ratios of $\Lambda_c^\prime\to \Sigma_c^\ast \pi$ is at least 100 times larger than the one of $\Lambda_c\to \Sigma_c^\ast \pi$, which means $\Sigma_c^\ast\pi$ could be a good channel for determining the $J^P$ quantum numbers of $\Lambda_c(2940)$ and $\Lambda_c(2910)$.

\subsection{The effect of the tensor coupling term in $\rho NN$ vertex}

In the aforementioned estimations, the tensor coupling terms in the $\mathcal{B} \mathcal{B}\mathcal{P}$ interactions, as depicted in Eq.~\eqref{Eq:Lag-Tensor}, were not taken into account. However, it is known that the tensor coupling plays a crucial role for the $\rho NN$ interaction~\cite{Yamaguchi:2019seo}, which potentially makes substantial contributions to the $ND$ channel. Therefore, the impact of the tensor coupling terms is further elaborated in this subsection. 

The detailed estimations are outlined in Appendix~\ref{Sec:App-B}. The estimated partial decay widths of $\Lambda_c(2910)$ and $\Lambda_c(2940)$ in scenario A are presented in Figs.~\ref{Fig:L2910-AT} and \ref{Fig:L2940-AT}. It is evident from the figures that the widths of the $ND$ channel are augmented as expected for both $\Lambda_c(2940)$ and $\Lambda_c(2910)$. Even including the tensor coupling term, the widths of $\Lambda_c(2940)$ and $\Lambda_c(2910)$ can still be simultaneously reproduced within a common $\Lambda$ range, which is $0.73\sim 0.77$. In Table~\ref{Tab:Br2}, the branching fractions of $ND$, $\Sigma_c \pi$ and $\Sigma_c^\ast \pi$ channels, determined by the common $\Lambda$ range in scenario A, are collected. For $\Lambda_c(2910)$, it dominantly decays into $\Sigma_c \pi $ and $ND$, with the branching fractions ratio approximately $3/2$. For $\Lambda_c(2940)$, it dominantly decays into $\Sigma_c^\ast \pi$, while the branching fractions of $\Sigma \pi$ and $ND$ channels are also sizable, and the ratios of the branching fractions of these channels are approximately predicted to be $\Sigma_c^\ast \pi:\Sigma_c \pi:ND=7:1:1$. 

The estimated branching fractions of $\Lambda_c(2940)$ and $\Lambda_c(2910)$ in scenario B are depicted in Figs.~\ref{Fig:L2940-BT} and \ref{Fig:L2910-BT}. The widths of $\Lambda_c(2940)$ and $\Lambda_c(2910)$ can be reproduced within the ranges $0.65< \Lambda <0.70$ and $0.74< \Lambda <0.88$, respectively. This suggests that both widths cannot be reproduced within a common parameter range. For comparison, we also collected the  branching fractions estimated with individual parameter range in scenario B in Table~\ref{Tab:Br2}, where one can also find the branching fractions of the $ND$ channel are largely enhanced.

Despite the changes in the magnitudes of the branching fractions due to the inclusion of the tensor coupling term in the $\rho NN$ vertex, the qualitative conclusions remain unchanged, i.e., scenario A is weakly favored and the $\Sigma_c^\ast \pi$ channel could be used to determined the $J^P$ quantum numbers of $\Lambda_c(2910)$ and $\Lambda_c(2940)$.

\section{Summary}
\label{Sec:summary}
The observed masses of $\Lambda_c(2940)$ and $\Lambda_c(2910)$ are a bit below the threshold of $ND^\ast$, and their mass splitting is similar to the one of $P_c(4440)$ and $P_c(4457)$; thus we consider both $\Lambda_c(2940)$ and $\Lambda_c(2910)$ as molecular states composed of $ND^{*}$ with different $J^P$ quantum numbers. Two possible combinations of $J^{P}$ quantum numbers are discussed in the present work. Scenario A corresponds to assuming that the $J^P$ quantum numbers of $\Lambda_c(2910)$ and $\Lambda_c(2940)$ are $1/2^-$ and $3/2^-$, respectively, while scenario B corresponds to the opposite identifications. 

The decay properties of $\Lambda_c(2910)$ and $\Lambda_c(2940)$ were investigated by using an effective Lagrangian approach in both scenarios, the widths of $ND$, $\Sigma_c \pi$ and $\Sigma_c^\ast \pi$ channels are estimated, and our results indicate that scenario A is weakly favored. In addition, we find the branching fractions of $\Lambda_c^\prime \to \Sigma_c^\ast \pi$ is at least 100 times larger than the one of $\Lambda_c\to \Sigma_c \pi$ in both scenarios. Thus only the signal of $\Lambda_c(3/2^-)$ is expected in the $\Sigma_c^\ast \pi$ invariant mass distributions, which indicates that the $\Sigma_c^\ast \pi$ channel could be a good channel to determine the $J^P$ quantum numbers of $\Lambda_c(2940)$ or $\Lambda_c(2910)$. Thus, we suggest the Belle II Collaboration search for the structure around 2900 MeV in the $\Sigma_c^\ast \pi$ invariant mass spectrum.

Before the end of this work, it is worth mentioning that, in Ref.~\cite{Belle:2006xni}, the Belle Collaboration observed $\Lambda_c(2940)$ in the $\Lambda_c^+ \pi^+ \pi^- $ invariant mass distributions, where both $\Lambda_c(2880)$ and $\Lambda_c(2940)$ were observed. In the $\Lambda_c(2880)$ yield as a function of $M(\Lambda_c^+ \pi^{\pm})$, there is a clear signal of $\Sigma_c (2455)$, and an excess of events in the region of $\Sigma_c(2520)$. As for the $\Lambda_c(2940)$ yield as the function of $M(\Lambda_c^+ \pi^{\pm})$, a clear signal of $\Sigma_c(2520)$ is expected if one considers the $J^P$ quantum numbers of $\Lambda_c(2940)$ as $3/2^-$, which could be tested by further measurements in Belle II.

\bigskip
\noindent
\begin{center}
  {\bf ACKNOWLEDGEMENTS}\\
\end{center}
The authors would like to thank Professor Yu-Bing Dong for useful discussions. This work is supported by the National Natural Science Foundation of China under the Grants No. 12175037 and No. 12335001.

\appendix
\section{THE  EFFECTIVE LAGRANGIAN WITH $\mathrm{SU(4)}$ SYMMETRY}
\label{Sec:App-A}

In $\mathrm{SU}(4)$ symmetry, the matrix form of pseudoscalar $\mathcal{P}$ and vector $\mathcal{V}$ meson 15-plet states are

\begin{eqnarray}
\mathcal{P}=
\begin{pmatrix}
\frac{\pi^{0}}{\sqrt{2}}+\frac{\eta}{\sqrt{6}}+\frac{\eta_{c}}{\sqrt{12}}&\pi^{+}&K^{+}&\bar{D}^{0}\\
\pi^{-}&-\frac{\pi^{0}}{\sqrt{2}}+\frac{\eta}{\sqrt{6}}+\frac{\eta_{c}}{\sqrt{12}}&K^{0}&D^{-}\\
K^{-}&\bar{K}^{0}&-\sqrt{\frac{2}{3}}\eta+\frac{\eta_{c}}{\sqrt{12}}&D_{s}^{-}\\
D^{0}&D^{+}&D_{s}^{+}&\frac{-3\eta_{c}}{\sqrt{12}}\nonumber\\
\end{pmatrix},
\end{eqnarray}
\begin{eqnarray}
\mathcal{V}=
\begin{pmatrix}
\frac{\rho_{0}}{\sqrt{2}}+\frac{\omega^{\prime}}{\sqrt{6}}+\frac{J/\psi}{\sqrt{12}}&\rho^{+}&K^{*+}&\bar{D}^{*0}\\
\rho^{-}&-\frac{\rho_{0}}{\sqrt{2}}+\frac{\omega^{\prime}}{\sqrt{6}}+\frac{J/\psi}{\sqrt{12}}&K^{*0}&D^{*-}\\
K^{*-}&\bar{K}^{*0}&-\sqrt{\frac{2}{3}}\omega^{\prime}+\frac{J/\psi}{\sqrt{12}}&D_{s}^{*-}\\
D^{*0}&D^{*+}&D_{s}^{*+}&-\frac{3J/\psi}{\sqrt{12}}\nonumber
\end{pmatrix},
\end{eqnarray}
respectively.


The baryon 20-plet states with $J^{P}=1/2^{+}$, $\mathcal{B}$, corresponding to $\mathrm{SU}(3)$ octet, can be denoted as
\begin{eqnarray}
p&=&\phi_{112}, ~~~~n=\phi_{221},~~~~\Lambda=\sqrt{\frac{2}{3}}(\phi_{321}-\phi_{312}),\nonumber\\
\Sigma^{+}&=&\phi_{113},~~~~\Sigma^{0}=\sqrt{2}\phi_{123},~~~~\Sigma^{-}=\phi_{223},\nonumber\\
\Xi^{0}&=&\phi_{331},~~~~\Xi^{-}=\phi_{332},\nonumber\\
\Sigma_{c}^{++}&=&\phi_{114},~~~~\Sigma_{c}^{+}=\sqrt{2}\phi_{124},~~~~\Sigma_{c}^{0}=\phi_{224},\nonumber\\
\Xi_{c}^{+}&=&\sqrt{2}\phi_{134},~~~~\Xi_{c}^{0}=\sqrt{2}\phi_{234},\nonumber\\
\Xi_{c}^{+\prime}&=&\sqrt{\frac{2}{3}}(\phi_{413}-\phi_{431}),~~~~\Xi_{c}^{0\prime}=\sqrt{\frac{2}{3}}(\phi_{423}-\phi_{432}),\nonumber\\
\Lambda_{c}^{+}&=&\sqrt{\frac{2}{3}}(\phi_{421}-\phi_{412}),~~~~\Omega_{c}^{0}=\phi_{334},\nonumber\\
\Xi_{cc}^{++}&=&\phi_{441},~~~~\Xi_{cc}^{+}=\phi_{442},~~~~\Omega_{cc}^{+}=\phi_{443}.
\end{eqnarray}

The baryon $J^{P}=\frac{3}{2}^{+}$ 20-plet states, $\mathcal{D}$, corresponding to decuplet in $\mathrm{SU}(3)$ symmetry, can be written as
\begin{eqnarray}
\Omega^{-}&=&D^{333},~~~~\Xi^{*0}=\sqrt{3}D^{133},~~~~\Xi^{*-}=\sqrt{3}D^{233},\nonumber\\
\Sigma^{*+}&=&\sqrt{3}D^{311},~~~~\Sigma^{*0}=\sqrt{6}D^{312},~~~~\Sigma^{*-}=\sqrt{3}D^{322},\nonumber\\
\Delta^{++}&=&D^{111},~~~~\Delta^{+}=\sqrt{3}D^{112},~~~~\Delta^{0}=\sqrt{3}D^{122},\nonumber\\
\Delta^{-}&=&D^{222},\nonumber
\end{eqnarray}
\begin{eqnarray}
\Sigma_{c}^{0}&=&\sqrt{3}D^{224},~~~~\Xi_{c}^{0}=\sqrt{6}D^{234},~~~~\Omega_{c}^{0}=\sqrt{3}D^{334},\nonumber\\
\Xi_{c}^{+}&=&\sqrt{6}D^{134},~~~~\Sigma_{c}^{++}=\sqrt{3}D^{114},~~~~\Sigma_{c}^{+}=\sqrt{6}D^{124},\nonumber\\
\Xi_{cc}^{+}&=&\sqrt{3}D^{244},~~~~\Xi_{cc}^{++}=\sqrt{3}D^{144},~~~~\Omega_{cc}^{+}=\sqrt{3}D^{344},\nonumber\\
\Omega_{ccc}^{++}&=&D^{444},
\end{eqnarray}
where $\phi_{ijk}$ and $D^{ijk}$ satisfy the following conditions, and $i$, $j$, $k$ can go from $1$ to $4$,
\begin{eqnarray}
\phi_{ijk}&+&\phi_{jki}+\phi_{kij}=0,\nonumber\\
\phi_{ijk}&=&\phi_{jik},\nonumber\\
D^{ijk}&=&D^{ikj}=D^{jki}=D^{jik}=D^{kij}=D^{kji}.
\end{eqnarray}

\begin{figure}[t]
\centering
  \includegraphics[width=8.5cm]{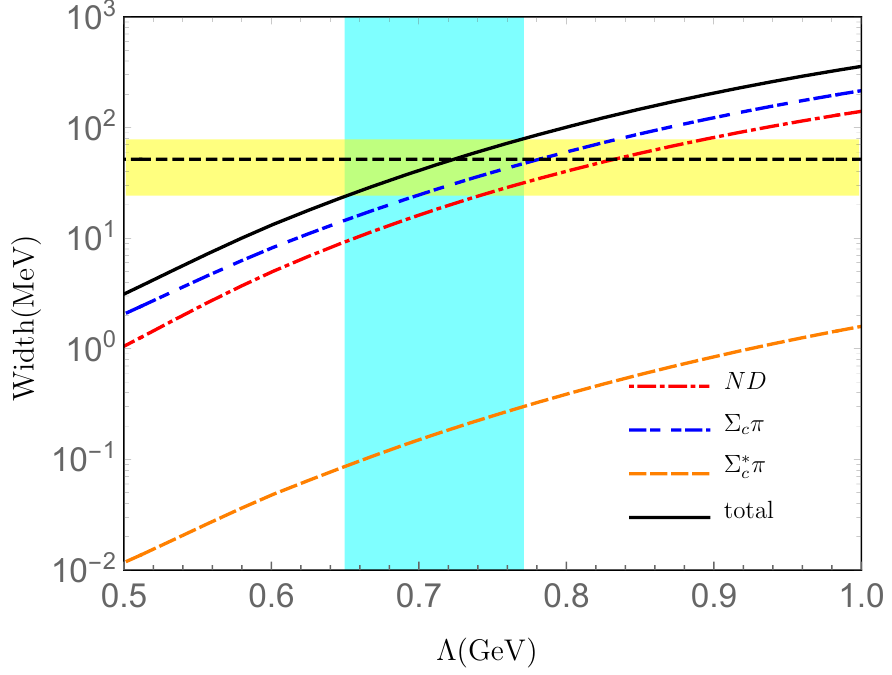}
\caption{The partial widths of $\Lambda_{c}(2910)\to ND$, $\Sigma_{c}\pi$, and $\Sigma_c^\ast \pi$ depending on the parameter $\Lambda$, where $\Lambda_c(2910)$ is assigned as $D^\ast N$ molecular state with $J^{p}=1/2^{-}$. }\label{Fig:L2910-AT}
\end{figure}

\begin{figure}[t]
  \centering
  \includegraphics[width=8.5 cm]{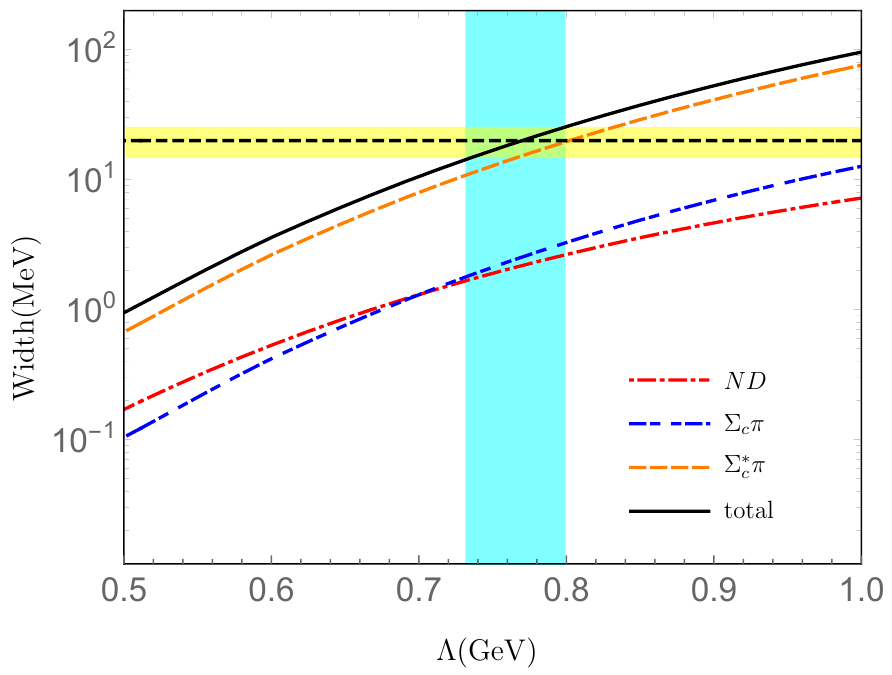}
\caption{The partial widths of $\Lambda_{c}(2940)\to ND$, $\Sigma_{c}\pi$, and $\Sigma_c^\ast \pi$ depending on the parameter $\Lambda$, where $\Lambda_c(2940)$ is assigned as $D^\ast N$ molecular state with $J^{p}=3/2^{-}$. }\label{Fig:L2940-AT}
\end{figure}

The $\mathrm{SU}(4)$ invariant effective Lagrangian related to baryons in tensor notions could be written as
\begin{eqnarray}
\mathcal{L}_{\mathcal{BBP}}&=&g_{p}(a\phi^{*ijk}\gamma_{5}\mathcal{P}_{i}^{m}\phi_{mjk}+b\phi^{*ijk}\gamma_{5}\mathcal{P}_{i}^{m}\phi_{mkj}),\nonumber\\
\mathcal{L}_{\mathcal{BBV}}&=&g_{v}(c\phi^{*ijk}\gamma^{\mu}\mathcal{V}_{i\mu}^{m}\phi_{mjk}+d\phi^{*ijk}\gamma^{\mu}\mathcal{V}_{i\mu}^{m}\phi_{mkj}),\nonumber\\
\mathcal{L}_{\mathcal{BDP}}&=&g_{p1}(e\phi^{*ijk}\partial_{\mu}\mathcal{P}_{i}^{m}D_{mjk}^{\mu}+f\phi^{*jki}\partial_{\mu}\mathcal{P}_{i}^{m}D^{\mu}_{mjk}),\nonumber\\
\mathcal{L}_{\mathcal{BDV}}&=&g_{v1}(g\phi^{*ijk}D_{mjk}^{\mu}+h\phi^{*kji}D_{mjk}^{\mu})\gamma^{5}\gamma^{\nu}\nonumber\\&\times&(\partial_{\mu}\mathcal{V}_{i\nu}^{m }-\partial_{\nu}\mathcal{V}_{i\mu}^{m}),\nonumber\\
\mathcal{L}_{\mathcal{DDP}}&=&g_{p2}(D_{\mu}^{*ijk}\gamma^{5}\gamma^{\nu}D_{mjk}^{\mu}\partial_{\nu}\mathcal{P}_{i}^{m}),\nonumber\\
\mathcal{L}_{\mathcal{DDV}}&=&g_{v2}(D^{*ijk}\gamma^{\mu}\mathcal{V}_{i\mu}^{m}D_{mjk}),
\end{eqnarray}
where $g_{p}$, $g_{v}$, $a$, $b$, $c$... are the constants associated with the couplings in the effective Lagrangian, which can be derived by comparing with the $\mathrm{SU}(3)$ relations\cite{Liu:2001yx,Ronchen:2012eg,Liu:2001ce}.


\begin{figure}[htb]
  \centering
\includegraphics[width=8.5 cm]{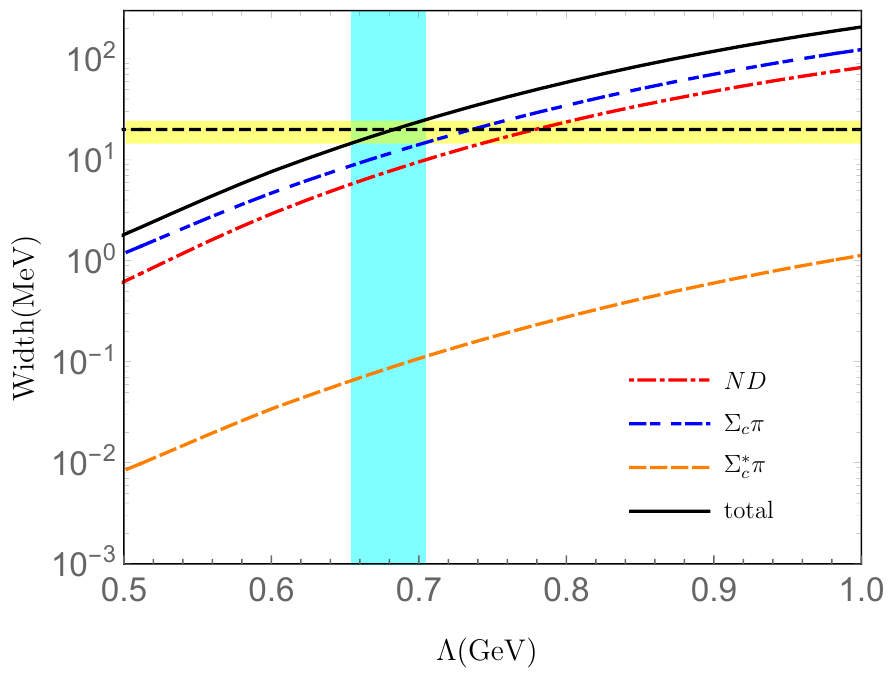}
\caption{The partial widths of $\Lambda_{c}(2940)\to ND$, $\Sigma_{c}\pi$, and $\Sigma_c^\ast \pi$ depending on the parameter $\Lambda$, where $\Lambda_c(2940)$ is assigned as $D^\ast N$ molecular state with $J^{p}=1/2^{-}$. }\label{Fig:L2940-BT}
\end{figure}

\begin{figure}[htb]
  \centering
\includegraphics[width=8.5 cm]{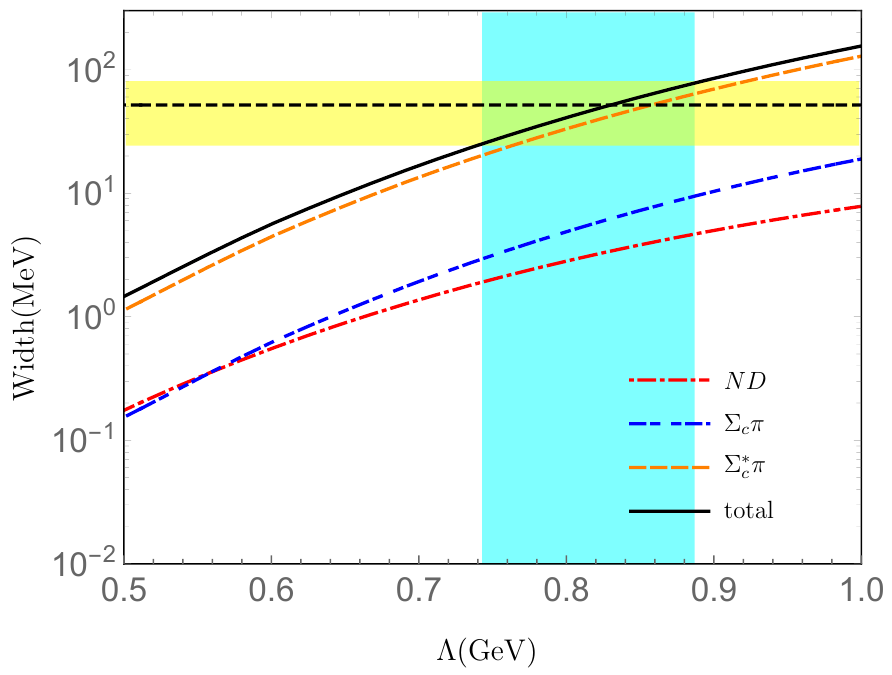}
\caption{The partial widths of $\Lambda_{c}(2910)\to ND$, $\Sigma_{c}\pi$, and $\Sigma_c^\ast \pi$ depending on the parameter $\Lambda$, where $\Lambda_c(2910)$ is assigned as $D^\ast N$ molecular state with $J^{p}=3/2^{-}$. }\label{Fig:L2910-BT}
\end{figure}

\begin{table*}[htb]
\renewcommand\arraystretch{1.75}
\caption{The branching fractions for $ND$, $\Sigma_c \pi$ and $\Sigma_c^\ast \pi$ channels for $\Lambda_c(2910)$ and $\Lambda_c(2940)$ in different scenarios with the tensor force.  \label{Tab:Br2}}
\begin{tabular}{p{2cm}<\centering p{3cm}<\centering p{3cm}<\centering p{0.2cm}<\centering p{3cm}<\centering p{3cm}<\centering}
\toprule[1pt]
\multirow{2}*{Channel} & \multicolumn{2}{c}{Scenario A (common $\Lambda$)} &&\multicolumn{2}{c}{Scenario B (individual $\Lambda$)}\\
\cmidrule[1pt]{2-3}  \cmidrule[1pt]{5-6}
    & $\Lambda_c(2910)$ & $\Lambda_c(2940)$ && $\Lambda_c(2940)$ &$\Lambda_c(2910)$ \\
 \midrule[1pt]
 $ND$ &  $(39.7\sim 39.9)\% $ & $(10.9\sim 11.7)\%$  && $(39.4\sim 40.0)\%  $ & $(6.06 \sim 7.62)\% $\\
 $\Sigma_c \pi$ & $(59.7\sim 59.9)\%$ & $(12.5\sim 12.7)\%$ && $(59.5\sim 60.2)\%$  &$(11.7\sim 12.1)\%$\\
 $\Sigma_c^\ast \pi $ & $(3.72\sim 3.79)\times 10^{-3}$ & $(75.8\sim 76.4)\%$ && $(4.48\sim 4.53)\times 10^{-3}$ & $(80.7\sim 81.8)\%$\\
\bottomrule[1pt]
\end{tabular} 
\end{table*}

\section{ESTIMATED RESULTS WITH TENSOR COUPLING TERM}
\label{Sec:App-B}
As indicated in Ref.~\cite{Yamaguchi:2019seo}, the tensor coupling term is important for the $\rho NN$ interaction~\cite{Yamaguchi:2019seo}, and the effective Lagrangians of the $\rho NN$ with tensor term reads
\begin{eqnarray}
\mathcal{L}_{\rho NN}=g_{\rho NN}\bar{\psi}_{N}\left[\gamma_{\mu}+\frac{\kappa_{\rho}}{2m_{N}}\sigma_{\mu\nu}\partial^{\mu}\right]\vec{\rho}^{~\nu}\cdot\tau\psi_{N}
\label{Eq:Lag-Tensor}
\end{eqnarray}
where $g_{\rho NN}^{2}/4\pi=0.84$. As for the value of $\kappa_\rho$, it was fitted to be  $1.825$ or $2.2176$ by reproducing the $\pi N$ scattering and $\gamma N \to \pi N$ reaction data in Ref. \cite{Sato:1996gk}, while it was determined to be $6.1$ in the Bonn meson exchange model for the $NN$ interaction~\cite{Machleidt:1987hj}. Here, we take the larger value of $\kappa_\rho$, i.e., $\kappa_\rho=6.1$, to check the effect of the tensor coupling term. The estimated branching fractions of the relevant decay channels depending on the model parameter $\Lambda$ in both scenarios are presented in Figs.~\ref{Fig:L2910-AT}-\ref{Fig:L2910-BT}. In Table~\ref{Tab:Br2}, the branching fractions estimated in a common $\Lambda$ range in scenario A and in the individual $\Lambda$ range in scenario B are presented.


\end{document}